# Helium-Implantation-Induced Lattice Strains and Defects in Tungsten probed by X-ray Micro-diffraction

S. Das[a], W. Liu[b], R. Xu[b], F. Hofmann[a]*

[a]*Department of Engineering Science, University of Oxford, Parks Road, Oxford OX1 3PJ, UK*
[b]*Advanced Photon Source, Argonne National Lab, 9700 South Cass Avenue, Argonne, IL 60439, USA*

* correspondence should be addressed to *felix.hofmann@eng.ox.ac.uk*

## Abstract

Tungsten is the main candidate material for plasma-facing armour components in future fusion reactors. Bombardment with energetic fusion neutrons causes collision cascade damage and defect formation. Interaction of defects with helium, produced by transmutation and injected from the plasma, modifies defect retention and behaviour. Here we investigate the residual lattice strains caused by different doses of helium-ion-implantation into tungsten and tungsten-rhenium alloys. Energy and depth-resolved synchrotron X-ray micro-diffraction uniquely permits the measurement of lattice strain with sub-micron 3D spatial resolution and ~$10^{-4}$ strain sensitivity. Increase of helium dose from 300 appm to 3000 appm increases volumetric strain by only ~2.4 times, indicating that defect retention per injected helium atom is ~3 times higher at low helium doses. This suggests defect retention is not a simple function of implanted helium dose, but strongly depends on material composition and presence of impurities. Conversely, analysis of W-1wt% Re alloy samples and of different crystal orientations shows that both the presence of rhenium, and crystal orientation, have a comparatively small effect on defect retention. These insights are key for the design of armour components in future reactors where it will be essential to account for irradiation-induced dimensional change when predicting component lifetime and performance.

## Introduction

Materials able to withstand the extreme conditions inside future fusion reactors are one of the main challenges for the development of commercial fusion power. Plasma facing armour components will be exposed to high temperatures (above 800 $^0$C), high heat loads (~10-20 MW/m$^2$ [1]) and intense ion and neutron irradiation [1,2]. Numerous parameters must be considered when selecting armour materials; low-activation rate to limit radioactive waste production [3], and a suitable operational temperature window, where thermal load capacity, thermal conductivity and recrystallization govern the upper temperature limit, whilst the lower temperature limit is controlled by ductile to brittle transition and low temperature irradiation resistance [1]. Low tritium retention and good resistance to surface erosion under irradiation are also essential. Tungsten, with an operational window of 500-1200 $^0$C [4], high melting point of (3422 $^0$C), low tritium inventory, low sputtering rate and reasonable strength at high temperatures has emerged as the main candidate material for plasma facing fusion armour [5,6].

Inside the reactor, substantial changes of the properties of tungsten armour components are expected during service. Neutron irradiation causes atomic displacements, leading to the creation and accumulation of defects, and results in hardening, embrittlement and irradiation creep [7]. Nuclear reactions further lead to transmutation and subsequent build-up of rhenium (~0.2 at. % year$^{-1}$), tantalum (~5000 appm after five year exposure) and osmium (~0.1 at. % year$^{-1}$), as well as hydrogen and helium [7,8]. The material will also be exposed to intense hydrogen and helium bombardment from the plasma (~10 MW/m$^2$) [9]. High interstitial mobility of these gasses, particularly at temperatures above 1000$^0$C, allows them to migrate to defects and grain boundaries, where they accumulate and lead to embrittlement and swelling [9]. For a five year operational lifetime, neutron damage in the first wall is anticipated to cause damage of up to 120 displacements per atom (dpa), while the accumulated helium content may reach up to 1200 appm [10].

For commercially viable fusion energy, reliable estimates of the functional properties and integrity of armour components are essential. They require a fundamental understanding of the generation and retention of irradiation-induced defects, and of how they modify material properties. Here helium plays a pivotal role due to its low solubility in the crystal lattice and its strong affinity for lattice defects [7]. A key question is to what extent helium modifies defect retention and whether the presence of other alloying elements, such as rhenium, has a significant effect on this.

Helium ion implantation provides a convenient way of simultaneously mimicking damage formed by energetic neutron irradiation and introducing helium into the tungsten matrix so that its interaction with defects may be studied [11–13]. Helium-ion implantation induced recoils have predominantly low energy, meaning that the damage micro-structure is dominated by Frenkel pair formation [12]. Probing these defects is challenging as transmission electron microscopy (TEM) is not sufficiently sensitive to detect them [14]. This is highlighted by high resolution TEM images of tungsten samples implanted with 3000 appm helium at room temperature that do not show any visible defects [15]. Though small, helium-implantation-induced defects have a significant impact on mechanical properties [11,15] and thermal conductivity of tungsten [16]. To analyse these effects, detailed knowledge of the underlying defect population is required. These studies are further complicated by the limited penetration depth of ions, meaning that implanted layers are generally only a few microns thick.

Several authors have used positron annihilation spectroscopy (PAS) to study vacancy type point defects in crystalline materials such as uranium dioxide or silicon-carbide [17,18]. This non-invasive technique can effectively be used to ascertain the presence of open-volume defects in which the positrons get trapped altering the positron lifetime. Past studies for example have successfully traced the process of vacancy migration and evolution of vacancy clusters with changing temperature and presence of negatively charged non-vacancy defects generated by proton irradiation [18]. Quantification of the vacancy concentration is also obtainable by the technique. However, interpretation of the PAS measurements requires supportive experimental methods or characterization techniques such as electron paramagnetic resonance (EPR). Further complexities may include distinction of vacancies from nano-voids or identification of defect combinations of vacancies and self-interstitials.

An alternative approach to probing these implantation-induced "invisible defects" is to measure the lattice strains they cause. This can be done using synchrotron X-ray micro-diffraction which uniquely permits the determination of lattice strain with sub-micron 3D spatial resolution and ~10$^{-4}$ strain sensitivity [11,12,19,20]. Using a combination of this approach and electronic structure calculations, we previously found that implantation of 3000 appm of helium into tungsten produced a defect structure dominated by Frenkel defects [12]. Recombination of Frenkel pairs was prevented by helium occupied vacancies ($He_2V$), which instead formed a stable configuration where the self-interstitial remains closely bound to the $He_2V$ [21,22]. Based on the damage microstructure determined from lattice strains, changes in the elastic properties due to defects could then be predicted in very good agreement with experiments [12,23].

Irradiation-defect-induced lattice strains are also important because of the residual stresses they introduce, which can be on the order of several hundred MPa [12,24]. These irradiation-induced residual stresses will add to structural in-service loads, creating a stress-offset. It is vital that this is accounted for when estimating the fatigue performance of key components [25]. A key question when estimating the strains, and hence stresses, caused by irradiation damage concerns the number of defects caused by particular irradiation conditions. This is complicated by the fact that generally only a small proportion of the defects produced in collision cascades are retained [12,26,27] and the vast majority of defects recombine soon after generation. Kinetic Monte Carlo simulations have shown that the defect retention does not scale linearly with the incident dose and depends on a number of other factors, such as the presence of impurities [28].

In this study, we use micro-beam Laue diffraction to examine the lattice strains caused by different doses of helium-ion implantation. Tungsten and tungsten rhenium alloys are considered to shed light on the importance of alloying elements for irradiation damage retention. The effect of different grain-orientations is considered to assess whether irradiation-induced defects preferentially orient depending on the ion-irradiation direction.

# Experimental details

## 2.1 Sample preparation

Two materials are considered: Polycrystalline, nominally pure (99.99%) tungsten (W), and a polycrystalline tungsten-1wt.%rhenium alloy (W-1Re) manufactured by arc melting from high purity elemental powders [29]. Two samples, ~5 mm in size and ~1 mm thick, were cut from each material. The sample surfaces were mechanically ground and then polished using diamond paste. A final chemo-mechanical polishing step with 0.1 µm colloidal silica suspension was used to produce a high-quality surface finish. Although polycrystalline, the grain size in the samples was on the order of ~1 mm. Thus, as far as the implantation and micro-diffraction measurements are concerned, the samples essentially behave as single crystals. The lattice orientation of grains was determined by electron backscatter diffraction (EBSD).

## 2.2 Ion implantation

Part of each sample was implanted with helium ions using a 2 MeV ion accelerator at the National Ion Beam Centre, University of Surrey, UK. Implantation was carried out at a temperature of 573 K and using a raster scanned beam, to ensure a uniform implantation dose. A range of ion energies (0.05 to 1.8 MeV) and fluences was used to obtain a near uniform helium ion concentration within a ~2.8 µm thick implanted layer [9,12,29]. Two samples, W and W-1Re were implanted with ~3000 appm of helium, while the two other samples, W and W-1Re, were implanted with a lower dose of ~300 appm helium. Overall four samples were produced, from here on referred to as W-300He, W-1Re-300He, W-3000He and W-1Re-3000He. Details of the ion energies and fluences used are provided in Appendix A. Figure 1 shows the implantation profiles as estimated using the SRIM code [30] (displacement energy of 68 eV, single-layer calculation model [31]). For W-3000He, between 0 and 2.8 µm depth, a helium ion concentration of ~3110 ± 270 appm is obtained with an associated damage of 0.24 ± 0.02 displacements per atom (dpa). For W-300He, between 0 and 2.8 µm depth, a helium ion concentration of ~310 ± 30 appm is obtained with an associated damage of 0.02 ± 0.003 displacements per atom (dpa).

## 2.3 X-ray micro-diffraction measurements

To measure lattice swelling due to helium ion implantation, micro-beam Laue diffraction measurements were carried out at beamline 34-ID-E, Advanced Photon Source, Argonne National Lab, USA. A polychromatic X-ray beam (7-30 keV) was focused by KB mirrors to a probe spot of ~300 nm full width at half maximum at the sample. The sample was placed at this probe spot in 45° reflection geometry (Figure 2). Laue diffraction patterns were recorded on an area detector (Perkin-Elmer, #XRD 1621, pixel size 200 x 200 µm$^2$) placed ~511 mm above the sample. This instrument uniquely allows measurements of lattice orientation and strain with sub-micron resolution in 3D, using the Differential Aperture X-ray Microscopy (DAXM) technique. Briefly, in the DAXM technique a ~50 µm diameter platinum wire is scanned in small steps between the detector and the diffracting sample. The depth vs intensity profile for each pixel on the detector can then be calculated by subtracting the diffraction images from consecutive wire position increments and triangulating using the wire edge and the line of the incident beam. This allows the reconstruction of depth-resolved diffraction data and hence 3D-resolved strain measurements in crystalline samples. A detailed description of the DAXM technique [20] and the 34-ID-E instrument [32–34] is provided elsewhere.

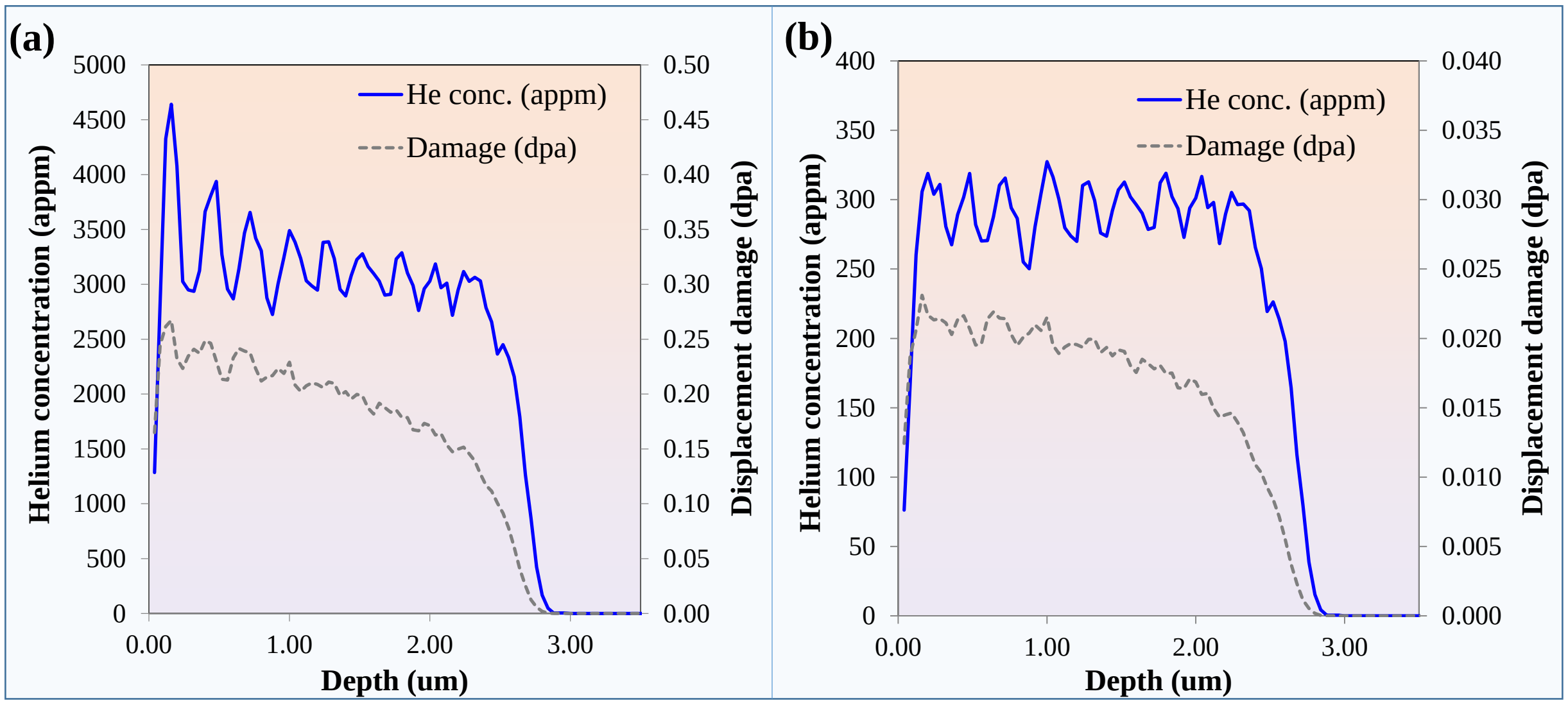

**Figure 1** - Profile of injected helium ion concentration (dotted grey curve) as calculated by SRIM and implantation-induced displacement damage (solid blue curve) as a function of depth in the sample.

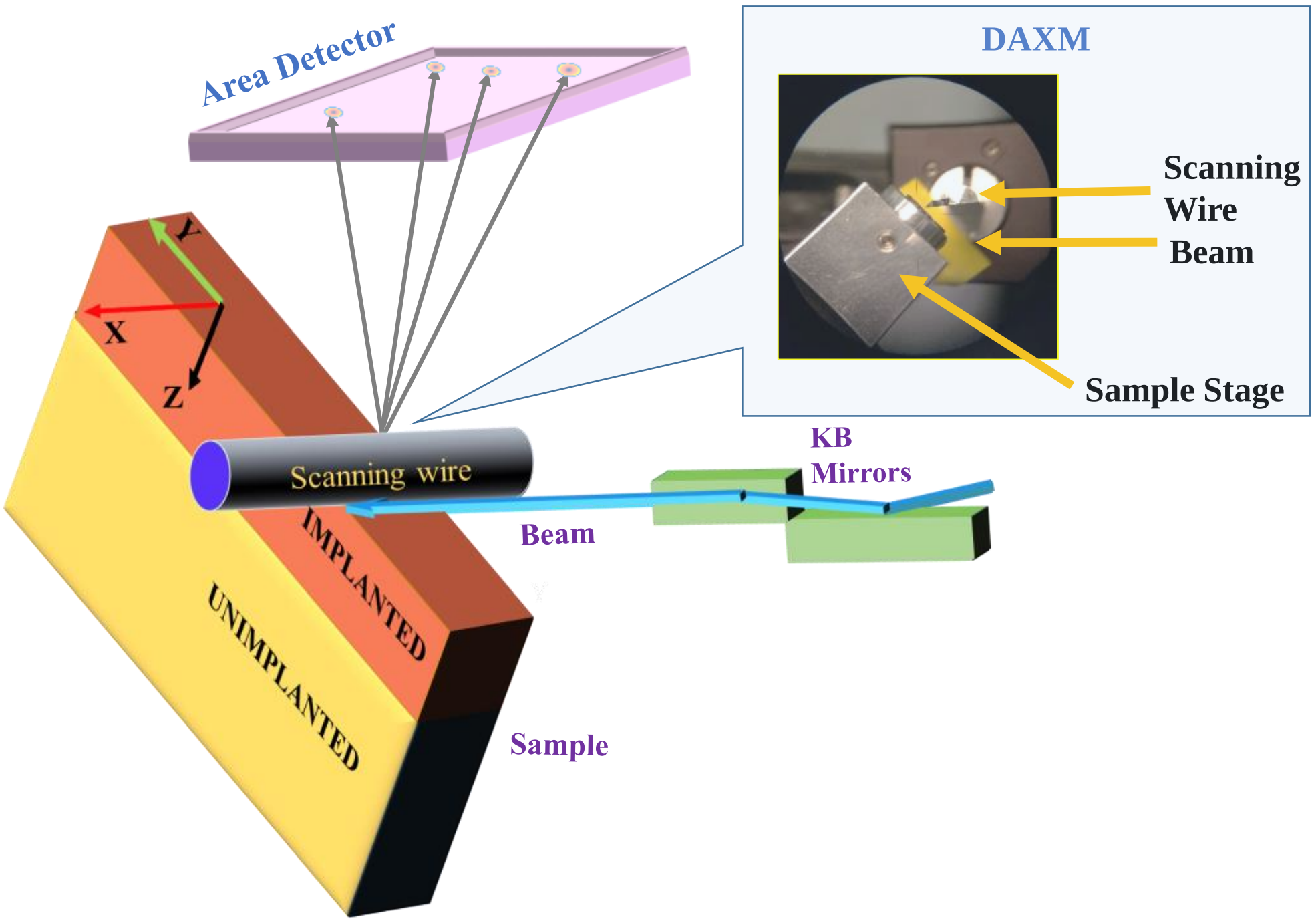

**Figure 2** - Laue diffraction measurements: Schematic of the experimental configuration illustrating the orientation of the sample with respect to the incident beam, scanning wire and detector, as well as the orientation of the sample coordinate axes.

## 2.4 Analysis of X-ray diffraction measurements

Polychromatic Laue measurements allow rapid measurement of the deviatoric lattice strain tensor only, as the radial position of diffraction peaks in reciprocal space is not known [19]. By scanning the energy of the incident X-ray beam, the full 3D reciprocal space position of specific reflections can be determined. This can be combined with the deviatoric strain tensor from a white beam measurement to find the full lattice strain tensor [12,35]. Alternatively the full lattice strain tensor can be determined by measuring the reciprocal space position of at least three non-collinear reflections [19,36]. This approach offers better sensitivity to small lattice strains and was followed here. For each measurement point photon energy- and depth-resolved scans were performed for six reflections, to ensure that the equations to compute the strain tensor are over determined. Reflections covering the widest possible angular range were chosen. For each reflection an energy range of ~80 eV was scanned with an energy step size of 2 eV, and a separate wire scan at each energy. This data was analysed and mapped into a 4 dimensional space (orthogonal reciprocal space $Q_x$, $Q_y$, $Q_z$, and distance along the beam direction) using the LaueGo software package (J.Z. Tischler: tischler@anl.gov).

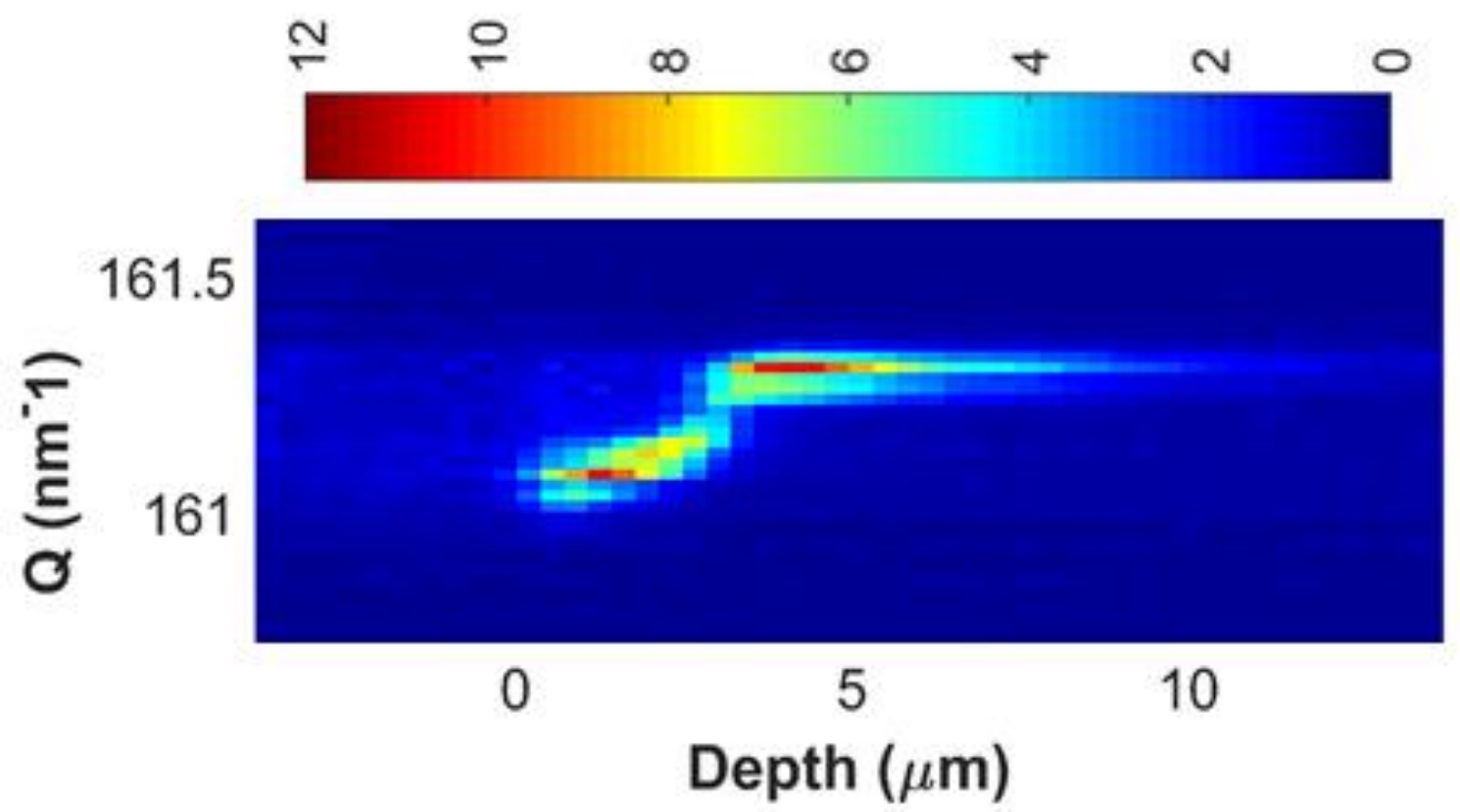

**Figure 3** – Integrated diffracted intensity on the area detector from a chosen sample point in a <111> oriented grain in the W-1Re-3000He sample as recorded on the detector. The measurement is for a (471) diffraction peak. The intensity is shown as a function of scattering vector magnitude |Q| and depth in the sample.

Figure 3 shows the integrated diffracted intensity of a (471) reflection recorded at a point in a <111> oriented grain in the W-1Re-3000He sample. The intensity is plotted as a function of scattering vector magnitude (Q) and depth in the sample. Two distinct peaks are visible: A broad peak at lower Q corresponding to the helium-implanted layer (0 to 2.7 µm depth), and a sharp peak from the unimplanted substrate (above 2.7 µm depth). The broadened peak associated with the implanted layer extending approximately 2.7 µm beneath the sample surface is in good agreement with the estimated implanted layer thickness predicted by SRIM (Figure 1) [30]. Similar agreement is observed for the other reflections measured in this and the other samples.

A detailed description of full strain tensor determination from three or more reflections with known 3D reciprocal space position can be found elsewhere [19]. Briefly, for each reflection, for each measured point along the depth of the sample, the intensity data was plotted in 3D reciprocal space and the peak centre was found using 3D Gaussian fitting and also confirmed using centre-of-mass. This was repeated for all six reflections at a particular measurement point. Knowing the peak centre for each reflection ($e.g. Qx_{refl1}, Qy_{refl1}, Qz_{refl1}$) and the $h,k,l$ indices for that reflection, the **UB** matrix was found, where **U** denotes the orientation matrix and **B** the matrix of reciprocal space lattice vectors.

$$\mathbf{UB}\begin{bmatrix} h_{refl1} & h_{refl2} & : & : & h_{refl7} \\ k_{refl1} & k_{refl2}: & : & : & k_{refl7} \\ l_{refl1} & l_{refl2}: & : & : & lh_{refl7} \end{bmatrix} = \begin{bmatrix} Qx_{refl1} & Qx_{refl2} & : & Qx_{refl7} \\ Qy_{refl1} & Qy_{refl2} & : & Qx_{refl7} \\ Qz_{refl1} & Qz_{refl2} & : & Qx_{refl7} \end{bmatrix} \quad (1)$$

If three non-collinear reflections are used, Eq. (1) can be solved directly. If more than three reflections are used Eq. (1) is overdetermined and the pseudo-inverse of the *hkl* matrix can be used to find the least squares solution for **UB** (here this was done using the mrdivide function in MATLAB).

From the **UB** matrix, the **U** and the **B** matrices can be found, using the information that $\mathbf{UU}^T = \mathbf{I}$, where **I** is the identity matrix and that the upper triangle of **B** contains zeros [19,37]. The real-space distorted lattice vectors **A** can then be found from **B**. Comparing **A** to the real-space lattice vectors of an undistorted tungsten crystal, $\mathbf{A}_0$, the deformation gradient, **G**, can be obtained as $\mathbf{A} = (\mathbf{G} + \mathbf{I})\,\mathbf{A}_0$. From **G** the strain tensor, $\boldsymbol{\varepsilon}_{ij}$ for the concerned measured point can be calculated as per Eq. (2).

$$\boldsymbol{\varepsilon}_{ij} = \frac{1}{2}\left(\mathbf{G}_{ij} + \mathbf{G}_{ji}\right) - \mathbf{I}_{ij} \tag{2}$$

The step is repeated for every measured point along the depth in the sample. To find $\mathbf{A}_0$ for the specific materials considered, we assume that the underlying, unimplanted substrate material is not affected by the ion-implantation and can be used as a built-in reference. Hence average data from depths of 7 to 10 µm beneath the sample surface was used to determine $\mathbf{A}_0$ for each material.

## 2.5 Samples chosen for analysis

To assess the influence of the grain orientation, two grains of approximate <111> and <001> surface normal orientation were selected in each sample for micro-diffraction measurements. Thus overall, eight points were studied, as shown by the black dots in Figure B.1 in the appendix. Table 1 shows the out-of-plane orientation of each chosen point and the misorientation of the corresponding point with respect to the perfect <111> or <001> out-of-plane direction.

| Sample | Euler Angles $(\varphi_1, \varphi, \varphi_2)$[1] | Out-of-plane orientation | Misorientation with <111> or <001> (in degrees) |
|---|---|---|---|
| **W-1Re-300He-<111>** | 332.2; 54.7; 319.4 | [53,62,58] | 4 |
| **W-1Re-300He-<001>** | 327.8; 18; 129.4 | [24,20,95] | 18 |
| **W-1Re-3000He-<111>** | 129.2;120.6;47 | [63,59,-51] | 5 |
| **W-1Re-3000He-<001>** | 341.4;87;354.1 | [-10,99,5] | 6.62 |
| **W-300He-<111>** | 225.3; 124.1; 222.4 | [-56,-61,-56] | 2.43 |
| **W-300He-<001>** | 7.5;76.3;196 | [-27, -93, 24] | 20.95 |
| **W-3000He-<111>** | 293.8; 55.2; 222 | [-55,-61,57] | 2.5 |
| **W-3000He-<001>** | 349.3; 14.6; 240.5 | [-22,-12,97] | 14.6 |

**Table 1** – List of the out-of-plane orientation of the chosen point in each sample grain and the misorientation of the chosen point with respect to the perfect <111> or <001> out-of-plane direction.

---

[1] The Euler angle convention used is as follows: Z1 = $\begin{bmatrix} \cos\varphi_1 & \sin\varphi_1 & 0 \\ -\sin\varphi_1 & \cos\varphi_1 & 0 \\ 0 & 0 & 1 \end{bmatrix}$; X = $\begin{bmatrix} 1 & 0 & 0 \\ 0 & \cos\varphi & \sin\varphi \\ 0 & -\sin\varphi & \cos\varphi \end{bmatrix}$; Z2 = $\begin{bmatrix} \cos\varphi_2 & \sin\varphi_2 & 0 \\ -\sin\varphi_2 & \cos\varphi_2 & 0 \\ 0 & 0 & 1 \end{bmatrix}$ and the rotation matrix R = Z1 * X * Z2.

## Results & Discussion

The Laue diffraction measurements performed on the helium-implanted sample measure the lattice strain, which is a combination of the eigenstrain induced by helium-implantation and the elastic strain generated in the sample in response to the eigenstrain.

The eigenstrain induced by helium can be realised by considering a small cube extracted from the helium-implanted tungsten layer, which is free to deform in all directions (traction free boundary conditions). The introduction of helium into the perfect crystal lattice of this tungsten cube will cause a volumetric expansion. We refer to the strain generated in the process as the volumetric eigenstrain, $\boldsymbol{\varepsilon}_{vol}^*$ (Figure D.1). Here, we make an assumption that due to random orientation of defects $\boldsymbol{\varepsilon}_{vol}^*$ is purely volumetric, i.e. $\boldsymbol{\varepsilon}_{vol}^* = \begin{bmatrix} \varepsilon_{\mathrm{vol}}^*/3 & 0 & 0 \\ 0 & \varepsilon_{\mathrm{vol}}^*/3 & 0 \\ 0 & 0 & \varepsilon_{\mathrm{vol}}^*/3 \end{bmatrix}$ [38]. With regard to the sample under study (a thick ~1 mm underlying tungsten substrate, with a thin helium-implanted layer ~ 2.8 µm sitting on top as shown in Figure D.1), this implies an expansive strain of $\varepsilon_{\mathrm{vol}}^*/3$ in the X and Y directions. However, the boundary interface between the implanted-layer and the substrate along the in-plane directions, X and Y (indicated by the dotted line AA' and A'B in Figure D.1), will prevent this expansion due to the need for geometrical continuity in the sample. Thus, an equal and opposite elastic compressive strain will be generated in the X and Y directions due to the presence of the boundary conditions. We refer to this elastic strain, generated in the sample in response to eigenstrain, as the correctional strain, represented by the tensor $\boldsymbol{\varepsilon}_{corr}$ with $\varepsilon_{corr_{11}} = \varepsilon_{corr_{22}} = -\varepsilon_{vol}^*/3$.

Eigenstrain when originating purely from plastic deformation (e.g. in the case of shot-peening [39]), remains undetected by diffraction measurements as there is no change in the average lattice spacing in the process of plastic deformation. However, the eigenstrain induced in the sample by helium-implantation, brings about a change in the lattice parameter and is thus measurable by the diffraction experiments. It is similar to a thermal strain in this respect. The experimentally measured lattice strain in the sample ($\boldsymbol{\varepsilon}$), is thus an additive combination of $\boldsymbol{\varepsilon}_{vol}^*$ and $\boldsymbol{\varepsilon}_{corr}$. The in-plane components of $\boldsymbol{\varepsilon}_{vol}^*$ and $\boldsymbol{\varepsilon}_{corr}$, being of equal magnitude and opposite sign cancel. Thus, the resultant in-plane strain components, as measured by the diffraction, are approximately equal to zero i.e. $\varepsilon_{xx} = \varepsilon_{yy} \approx 0$. This is consistent with our measurements, as seen in Fig. 4 and 5 and discussed below in further detail.

Figure 4 and Figure 5 show the average of the direct strains measured (averaged over 0-2.5 µm depth) in the helium-damaged layer of the samples implanted with 300 appm of helium and 3000 appm of helium respectively. The corresponding line plots over the whole depth range of 0-10 µm, are provided in Appendix C in Figure C.1 and Figure C.2 respectively. Figure 6 and Figure 7 show the average shear strains in the 2.5 µm implanted layer measured for samples implanted with 300 and 3000 appm of helium respectively. The corresponding line plots over the whole depth range of 0-10 µm, are provided in Appendix C in Figure C.3 and Figure C.4 respectively. The current experimental configuration and arrangement of the detectors is less sensitive to shear strains than direct strains [40]. Accordingly the experimental uncertainty associated with the reported shear strains are somewhat larger than those of the direct strains.

It is expected that lattice distortion will be close to zero in the X and the Y directions, due to the plane-strain condition imposed in these directions by the unimplanted substrate material [12,41]. Consequently, lattice strains brought about by implantation-induced defects, must be accommodated in the out-of-plane (z) direction. This can be clearly seen for all samples in Figure 4 and Figure 5, where the out-of-plane $\varepsilon_{zz}$, component is positive and significantly larger than the $\varepsilon_{xx}$ and $\varepsilon_{yy}$ components. These strains can be interpreted in terms of lattice swelling associated with helium implantation and lattice defects generated in the process, as discussed below. It is interesting to note that the W-Re sample shows substantially less fluctuation of strains with depth than the W samples. This is particularly noticeable in the line plots in Figure C.1 and Figure C.2. Strain fluctuation in the implanted layer are indicative of a heterogeneous defect population, suggesting that the presence of Re leads to a more homogeneous distribution of defects.

It is worth noting that the strains induced by the 300 appm helium implantation are rather small, on the order of a few $10^{-4}$. Measurements of such small strains were made possible here by the improved diffraction setup at 34-ID-E, with new, high stability KB mirror mounts [42,43] and a newly fabricated DAXM wire with reduced shape uncertainty.

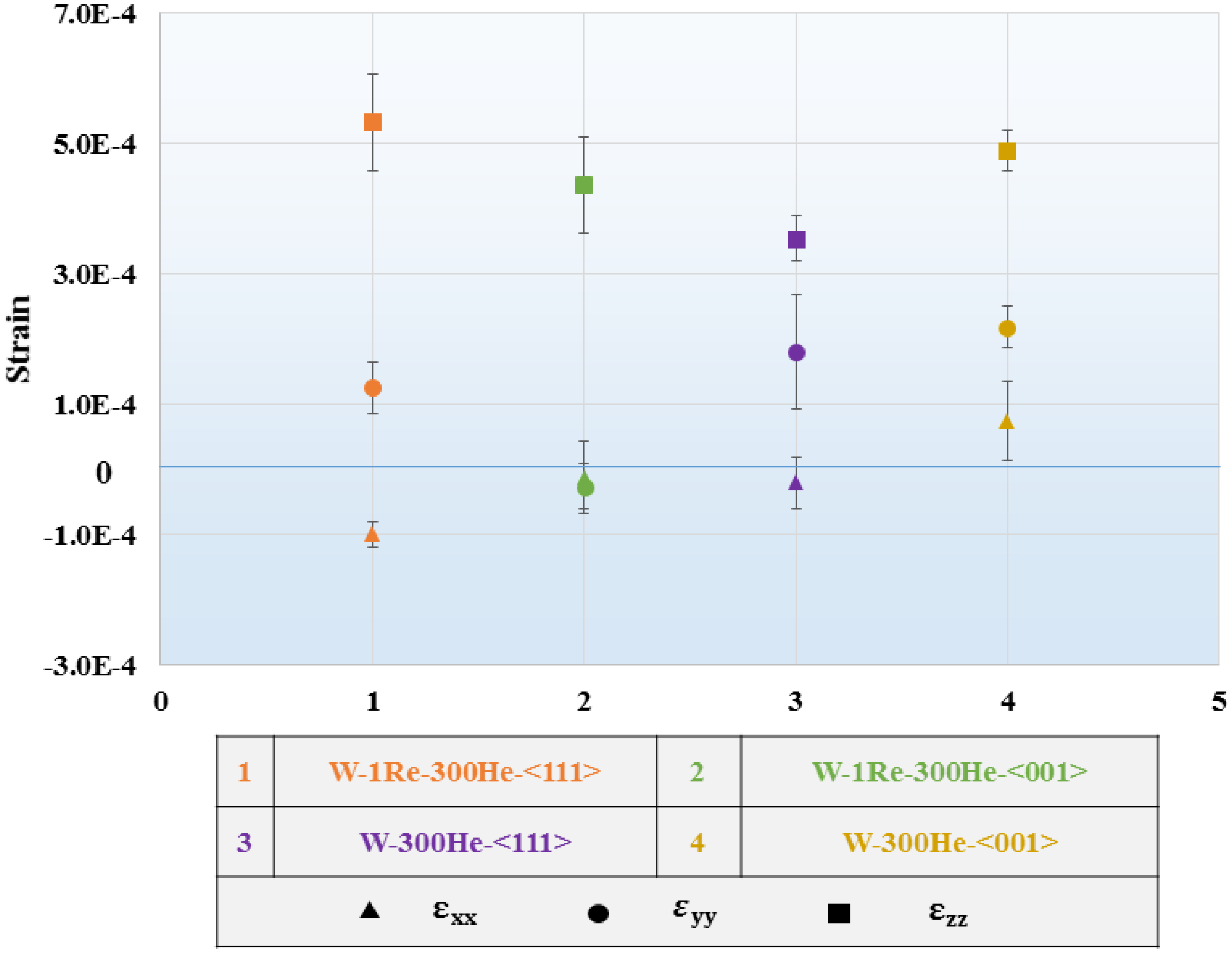

**Figure 4** – Average of the direct strains in the 2.5 µm helium-implanted layer in the samples implanted with 300 appm of helium. The error bars show ±1 standard deviation.

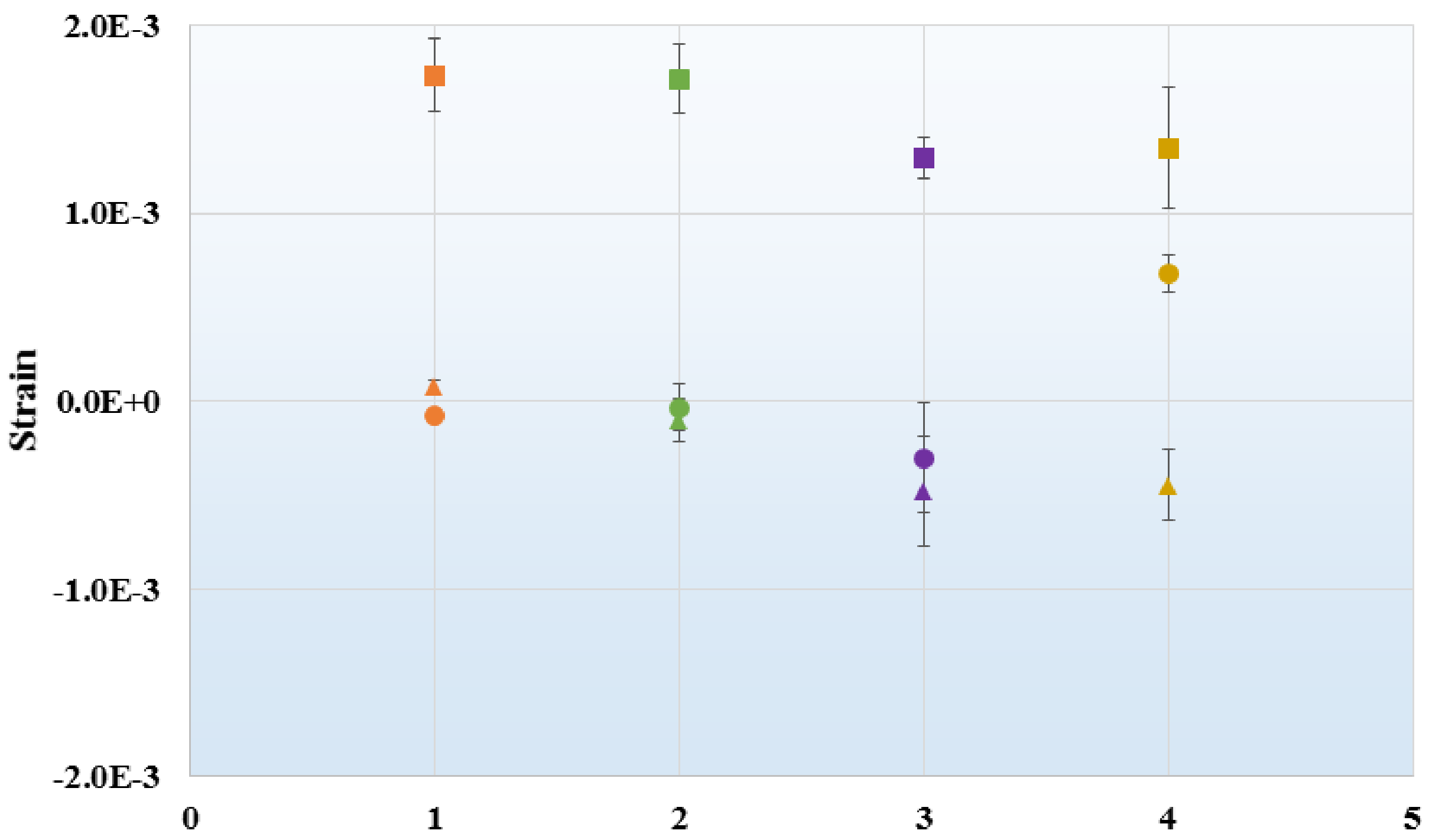

| | | | |
|---|---|---|---|
| 1 | W-1Re-3000He-<111> | 2 | W-1Re-3000He-<001> |
| 3 | W-3000He-<111> | 4 | W-3000He-<001> |
| ▲ $\varepsilon_{xx}$ | ● $\varepsilon_{yy}$ | ■ $\varepsilon_{zz}$ | |

**Figure 5** - Average of the direct strains in the 2.5 µm helium-implanted layer in the samples implanted with 3000 appm of helium. The error bars show ±1 standard deviation.

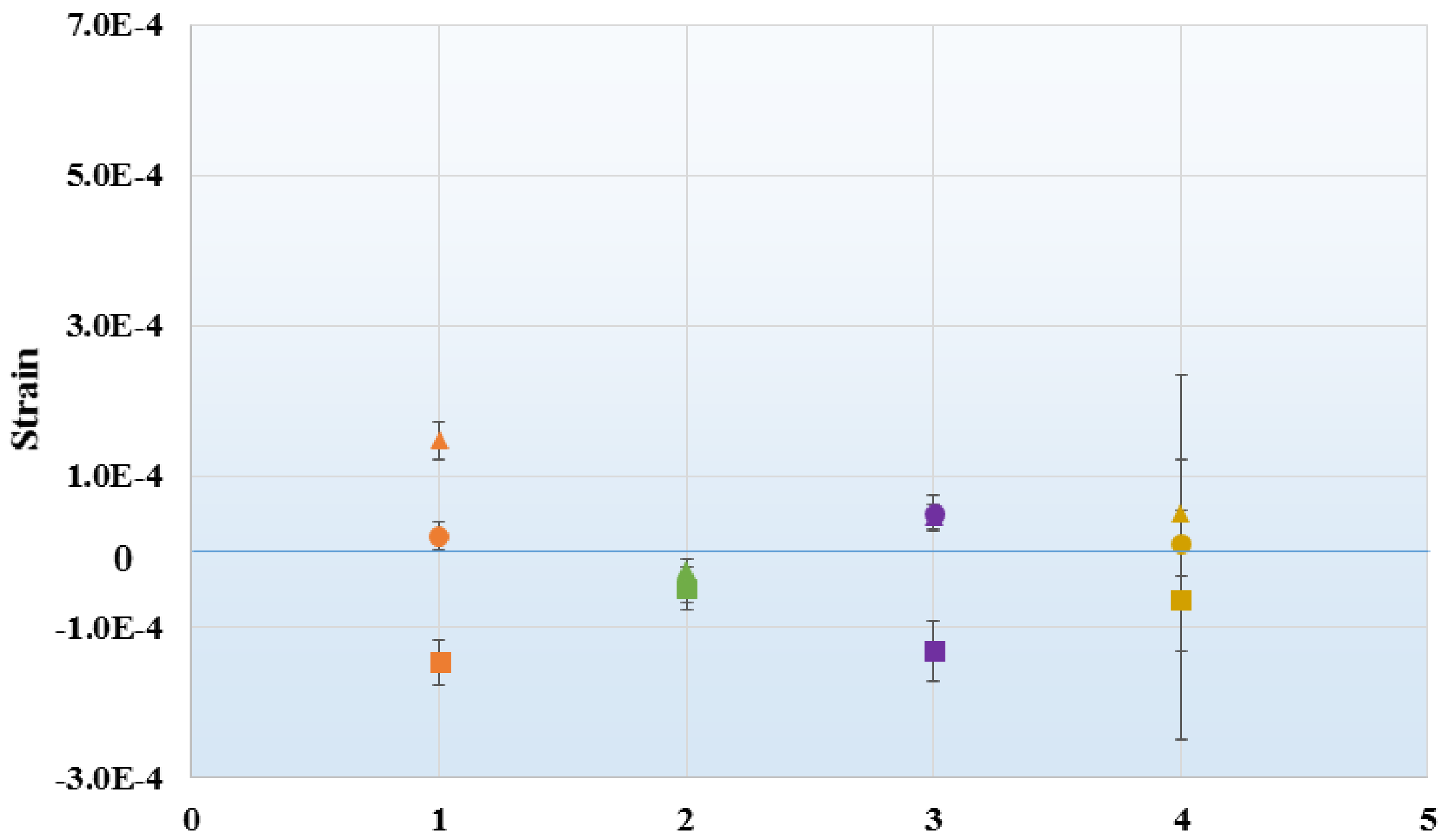

| | | | |
|---|---|---|---|
| 1 | W-1Re-300He-<111> | 2 | W-1Re-300He-<001> |
| 3 | W-300He-<111> | 4 | W-300He-<001> |
| ▲ $\varepsilon_{xy}$ | ● $\varepsilon_{xz}$ | ■ $\varepsilon_{yz}$ | |

**Figure 6** - Average of the shear strains in the 2.5 μm helium-implanted layer in the samples implanted with 300 appm of helium. The error bars show ±1 standard deviation.

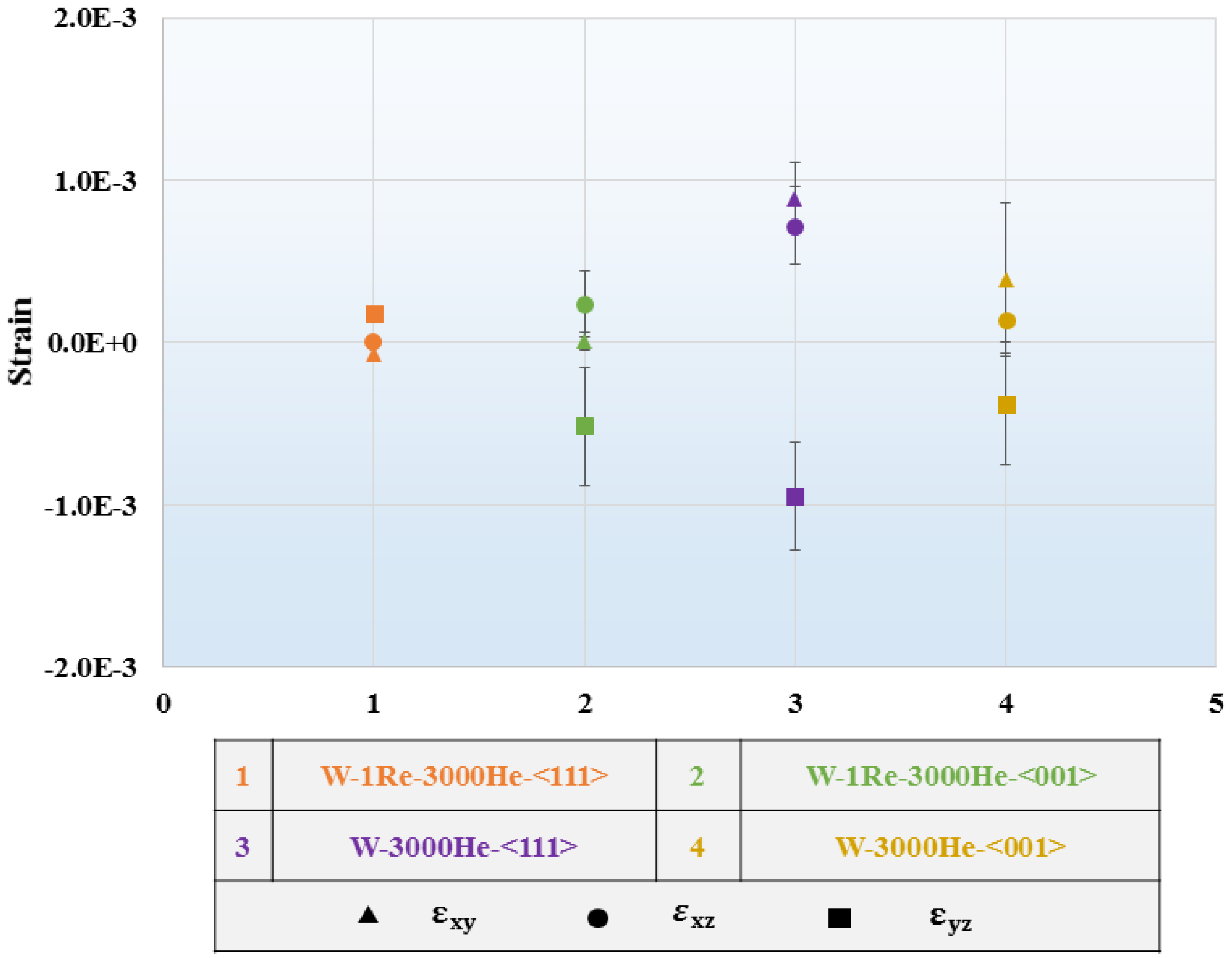

**Figure 7** - Average of the shear strains in the 2.5 µm helium-implanted layer in the samples implanted with 3000 appm of helium. The error bars show ±1 standard deviation.

At room temperature tungsten is almost perfectly elastically isotropic [44,45], and hence $\varepsilon_{zz}$ can be used to estimate the underlying Eigenstrain $\boldsymbol{\varepsilon}^*_{vol}$. $\varepsilon_{zz}$ can then be written in terms of the $\boldsymbol{\varepsilon}^*_{vol}$ as:

$$\varepsilon_{zz} = \frac{\varepsilon^*_{vol}}{3} + 2 * \left(\frac{\nu}{1-\nu}\right) * \left(\frac{\varepsilon^*_{vol}}{3}\right), \qquad (3)$$

where $\nu$ is Poisson's ratio, which for tungsten is 0.28 [44,45]. Details about the derivation of Eq. (3) are provided in Appendix D. The volumetric Eigenstrain for each of the samples, estimated using Eq. (3) and the measured $\varepsilon_{zz}$ strain, is listed in Table 2. Here, in each case, the average of $\varepsilon_{zz}$, measured between 0-2.5 µm was considered for the calculations. For the tungsten-rhenium alloy the same Poisson ratio as for pure tungsten ($\nu$ = 0.28) was assumed. We note that the addition of rhenium to tungsten [46] does not significantly alter the crystallographic properties of pure tungsten. The lattice constant measured in the tungsten-rhenium alloy samples, at depths greater than 7 µm below the sample surface, is 3.165±0.0009 Å. This is very close to the lattice constant for pure tungsten 3.16522±0.00009 Å [47] and to the value previously measured for material from the same batch of 3.165±0.0007 Å [12].

| Sample | Average of $\varepsilon_{zz}$ measured between 1-2 µm ($\times 10^{-3}$) | Calculated $\varepsilon^{*}_{vol}$ ($\times 10^{-3}$) |
|---|---|---|
| **W-1Re-300He-<111>** | 0.533 | 0.899 |
| **W-1Re-300He-<001>** | 0.436 | 0.736 |
| **W-300He-<111>** | 0.355 | 0.599 |
| **W-300He-<001>** | 0.489 | 0.825 |
| **W-1Re-3000He-<111>** | 1.74 | 2.934 |
| **W-1Re-3000He-<001>** | 1.72 | 2.896 |
| **W-3000He-<111>** | 1.30 | 2.188 |
| **W-3000He-<001>** | 1.35 | 2.272 |

**Table 2** – Volumetric Eigenstrain associated with He-implantation in each of the samples, calculated from the measured $\varepsilon_{zz}$ strain.

The effect of the three parameters varying between the samples can now be assessed: the helium dose, the presence of rhenium and the effect of grain orientation. We will deal with each aspect in turn. The average level of lattice strain for 300 and 3000 appm implanted samples (average taken over all four samples with 300/3000 appm helium) is $0.765\times 10^{-3}$ and $2.57\times 10^{-3}$ respectively. This corresponds to a ~240% increase in the underlying volumetric Eigenstrain for a ten-fold increase in the implanted helium dose. The volumetric strain is closely linked to the defects produced by the helium-ion-implantation. Our observation suggests that the number of defects retained per injected helium ion in the 300 appm implanted samples is three times greater than in the 3000 appm implanted samples. Interestingly, this is consistent with previous thermal transport measurements on 300 appm and 3000 appm helium-implanted tungsten samples [16]. Here too an approximately three times higher defect retention per helium ion was observed in the samples exposed to the lower helium dose.

An interesting question concerns the origin of this dose dependence of defect retention per injected helium ion. Calculations and previous experiments [12] suggest that the defects in the implanted layer take the form of vacancies, self-interstitials (SIAs) or small clusters of both [48]. SRIM calculations [30] indicate that, under the present implantation conditions, approximately 60 Frenkel pairs are formed for every injected helium ion. Most of these Frenkel pairs recombine almost immediately and only a small fraction, a few % (although the exact fraction depends on the implanted helium dose), are prevented from recombining by the presence of helium [12]. Since helium has a strong affinity for vacancies and vacancies are in plentiful supply during helium implantation, helium atoms become trapped at vacancies. Once the helium has occupied a vacancy, it prevents recombination of this vacancy with a self-interstitial atom, leading to the retention of some of the Frenkel pairs formed during implantation [21,22]. Becquart et al. [49] pointed out, however, that helium only plays a dominant role in defect retention at higher injected concentrations, i.e. when helium is the dominant impurity. For low helium concentrations other impurities present in the sample, most notably carbon, are primarily responsible for preventing the recombination of the Frenkel pairs. This suggests that, in a sample without any impurities, defect retention in the 300 appm helium-implanted samples and the associated volumetric strain should be much lower (~ estimated strain based on 3000 appm value divided by 10) than the volumetric strain we observe. Hence impurities clearly play an important role in controlling implantation-induced lattice strains, especially at low damage doses.

With regard to impurities, the tungsten-rhenium alloys investigated here have an additional impurity in the form of the rhenium atoms. The presence of rhenium is important since it is one of the elements that will build up in tungsten armour components due to transmutation, with an expected concentration of ~ 3.8 at. % after five years of operation [8]. Considering the four samples implanted with 300 appm of helium (rows 1-4 of Table 2), and comparing the average volumetric strain across samples with and without rhenium, it is seen that the presence of rhenium is associated with a ~15% increase in strain. A similar analysis on the four samples with 3000 appm of helium (rows 5-8 in Table 2) shows a ~30% increase associated with the addition of rhenium. Next, if we focus on the four tungsten-rhenium alloy samples (W-1Re-300-<111>/<001> & W-1Re-3000He-<111>/<001>), and monitor the change in volumetric strain with the tenfold increase in the helium dose, here a ~256% increase in $\varepsilon_{vol}^{*}$ is seen. This is slightly higher (relative percentage increase of ~35%) than the ~213%±5% increase in $\varepsilon_{vol}^{*}$ seen in the group of four pure tungsten samples (W-300-<111>/<001> & W-3000He-<111>/<001>) when the helium dose is changed from 300 to 3000 appm.

These observations suggest that, compared to helium (max. concentration 0.3 at. %), rhenium, even at ~1 wt. % concentration has relatively little effect on defect retention. From first principle studies [50] it has been seen that Re binds quite weakly with vacancies (binding energy of ~0.22 eV), but its binding energy with SIAs is quite high (0.8 eV). Similar estimations have been obtained by object Kinetic Monte Carlo simulations of point defects in tungsten by Becquart et al. [51] and from first principle calculations based on density functional theory (DFT) [52]. Although the Re and the SIA are tightly bound, this mixed-interstitial can migrate easily using a non-dissociative mechanism with a low migration barrier of 0.12 eV [52]. In other words, this implies that even in the presence of Re, the migration of SIAs is not strongly affected i.e. the presence of rhenium will not significantly alter defect retention. Further DFT calculations showed that addition of rhenium to tungsten may improve its ductility by changing dislocation core structure and thereby reducing the critical stress required for plasticity to set in [53]. This is consistent with nano-indentation observations where the presence of rhenium in tungsten had no significant effect on the hardening behaviour [54]. Rather, it has been seen through DFT that presence of rhenium in tungsten actually enhances vacancy-interstitial recombination (suggested that the vacancy suppression occurs at isolated interstitials) and suppression of void swelling, leading to the idea that addition of rhenium to tungsten can improve its radiation-resistance [55]. This of course must be balanced with irradiation-induced clustering of rhenium that sets in at high damage doses and leads to substantial hardening as rhenium clusters act as efficient obstacles to dislocation motion [56–58].

To assess the effect of lattice orientation, diffraction measurements were made in two grains in each sample, a <111> and a <001> oriented grain. In W-300He sample, the <001> grain shows ~38% higher strain than the <111> grain, while the corresponding percentage increase in W-3000He sample is ~4%. For the W-1Re-3000He sample no change in the volumetric strain is noticed between the grains. The W-Re-300He sample, on the other hand, shows a higher $\varepsilon_{vol}^{*}$ in the <111> grain compared to the <001> grain. These results, based on a small sample size, are insufficient to provide a definite answer to the question of how the grain orientation influences helium-implantation-induced lattice swelling. However, our results do suggest that the effects of grain orientation are most prominent for low helium concentrations and diminish with increasing the helium dose.

## Implications for Design

The findings from our experimental study have important implications for the design of future fusion reactors. The large dimensional change (lattice swelling) associated with helium-implantation damage in tungsten, and the associated stresses must be accounted for when estimating armour component fatigue life. In principal, this could be done using the framework recently proposed by Dudarev et al. [41].

A second important finding is the non-linear variation of defect retention as a function of injected helium dose. This effect must be properly characterised and models need to be developed for capturing this behaviour to allow reliable prediction of the evolution of the properties of tungsten armour. The reduced defect retention with increasing helium dose suggests that there may be a saturation regime, similar to observations in self-ion implantation damage in tungsten [59–61].

We also find that not all transmutation elements have a strong effect on defect retention. For design, one of the next key activities must be to map out the effect of different transmutation elements on defect retention as a function of concentration and damage level. This will require experimental effort, as well as computational models capturing the interaction of defects with transmutation elements. The results will directly feed into reactor armour design, enabling the estimation of defects produced and retained, and predictions of the resulting distortions.

Finally our results suggest that grain-orientation dependent effects become insignificant, particularly with increasing helium dose, implying that texture control may not be necessary in this regard. This is good news, as it will allow the optimisation of texture to enhance other aspects of fusion armour performance. For example, (111) texture may be used, as it has been seen to have substantially lower erosion yield than (001) oriented grains [62,63].

## Conclusions

In summary, our results show that energy- and depth-resolved micro-diffraction provides an effective means of probing lattice strains caused even by quite low doses of helium implantation into tungsten. Increasing the dose of helium from 300 appm to 3000 appm causes an increase of ~240% in volumetric Eigenstrain, indicating that the defect retention per injected helium atom is ~3 times higher at low helium doses. This highlights that implantation-induced lattice strains are not simply a linear function of dose, but follow a more complex relationship. Importantly defect retention is strongly depended on material composition and the presence of impurities. Analysis of W-1at.wt% Re alloy samples showed that the presence of rhenium has a relatively small effect on defect retention in tungsten. Similarly grain orientation appears to only lead to minor changes in the retained defect population. This is important as it means that, at least as far as implantation-induced strains are concerned, tight control of the texture of tungsten armour components may not be necessary.

## Acknowledgements

We thank Dr. C.E. Beck and Prof. D.E.J. Armstrong for providing the samples, and Dr. N. Peng for performing the ion-implantation. This work was funded by Leverhulme Trust Research Project Grant RPG-2016-190. This research used resources of the Advanced Photon Source, a U.S. Department of Energy (DOE) Office of Science User Facility operated for the DOE Office of Science by Argonne National Laboratory under Contract No. DE-AC02-06CH11357. Ion implantations were performed under the UK Engineering and Physical Sciences Research Council grant EP/H018921/1.

## Appendix A

List of 12 ion energies used and the corresponding fluence for the helium ion implantations:

| Ion Energy (MeV) | 3000 appm He | 300 appm He |
|---|---|---|
| | Fluence (ions/cm$^2$) | Fluence (ions/cm$^2$) |
| 0.05 | 3.60E+15 | 1.50E+14 |
| 0.1 | 1.50E+15 | 2.50E+14 |
| 0.2 | 1.00E+15 | 3.50E+14 |
| 0.3 | 5.00E+15 | 1.50E+14 |
| 0.4 | 5.00E+15 | 4.00E+14 |
| 0.6 | 5.00E+15 | 4.75E+14 |
| 0.8 | 5.00E+15 | 4.50E+14 |
| 1 | 5.00E+15 | 4.75E+14 |
| 1.2 | 5.00E+15 | 4.75E+14 |
| 1.4 | 5.00E+15 | 5.00E+14 |
| 1.6 | 5.50E+15 | 5.50E+14 |
| 1.8 | 6.00E+15 | 5.50E+14 |

**Table A.1** – List of 12 ion energies used and the corresponding fluence for the helium ion implantations.

## Appendix B

Micro-diffraction measurements were carried out in each of the pre-determined grains of each sample. Overall, eight points were investigated experimentally, as shown by the black dots in Figure B.1.

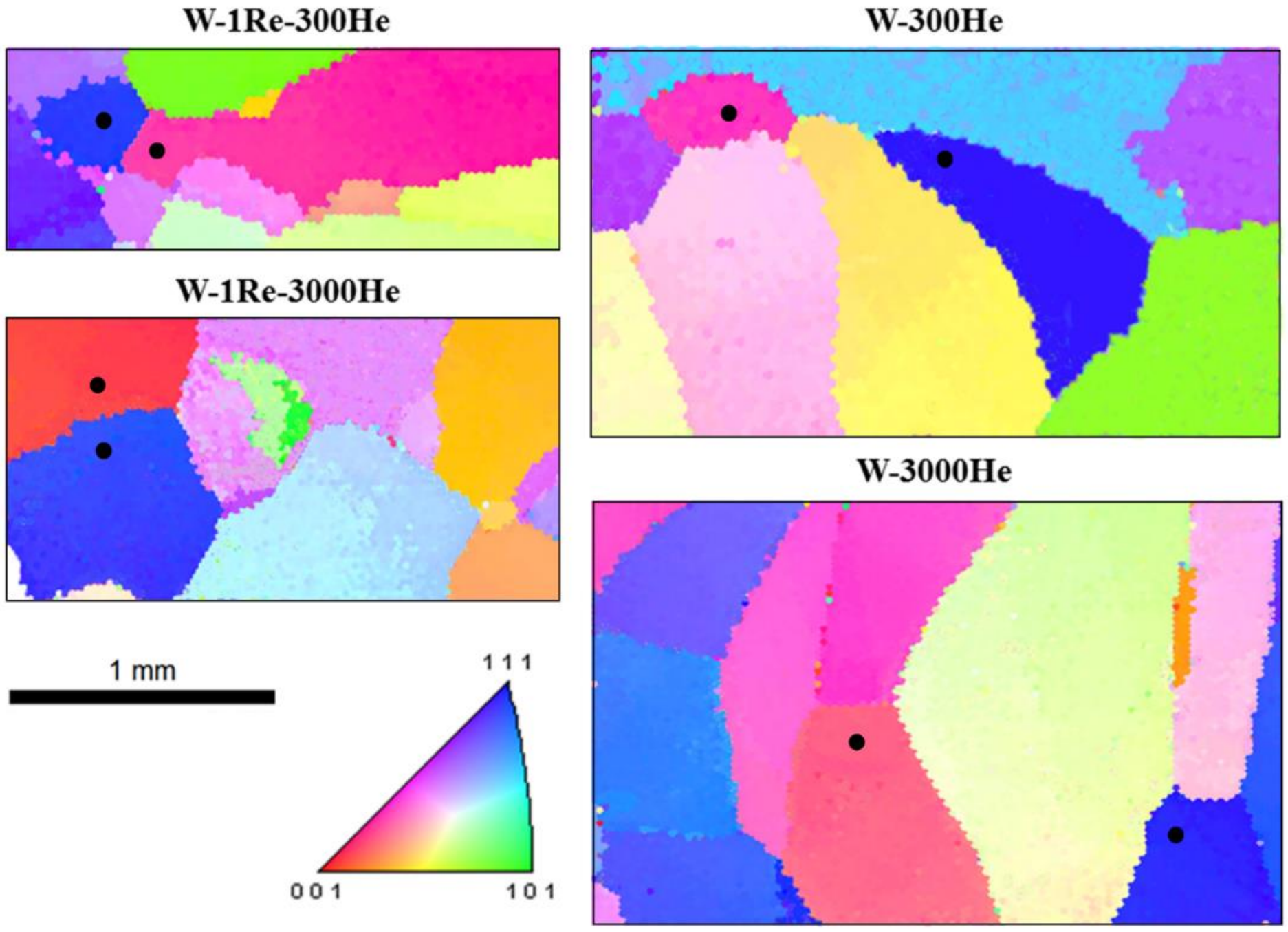

**Figure B.1** - EBSD images of the four samples used in the study. The black spots indicate the positions in each sample where the micro-diffraction measurement was carried out.

## Appendix C

Figure C.1 and Figure C.2 show the line plots for the direct strains over the depth range of 0-10 µm for the 300 and 3000 appm helium-implanted samples respectively. Figure C.3 and Figure C.4 show the line plots for the shear strains over the depth range of 0-10 µm for the 300 and 3000 appm helium-implanted samples respectively.

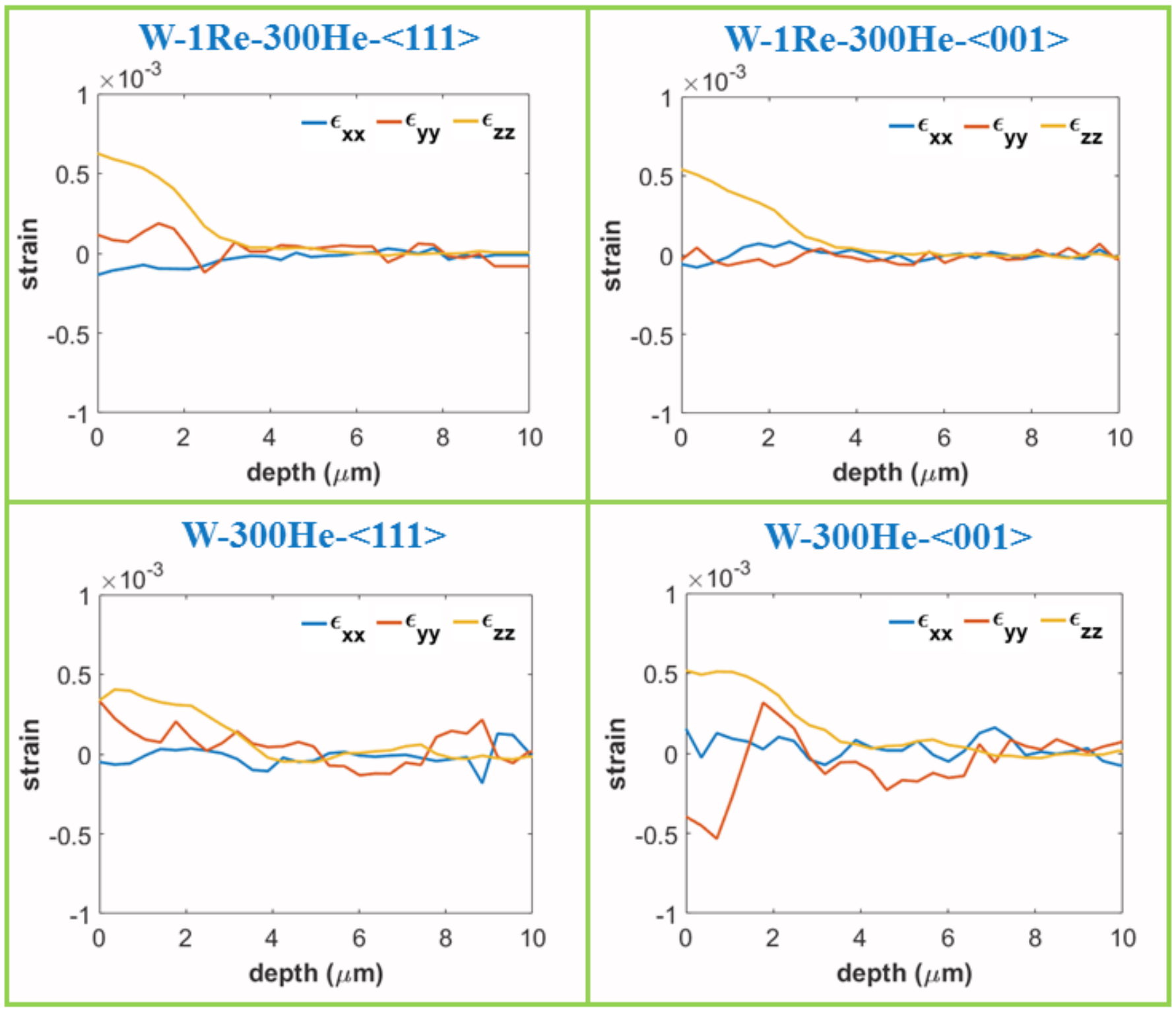

**Figure C.1** – Direct strains measured for samples implanted with 300 appm of helium plotted vs sample depth.

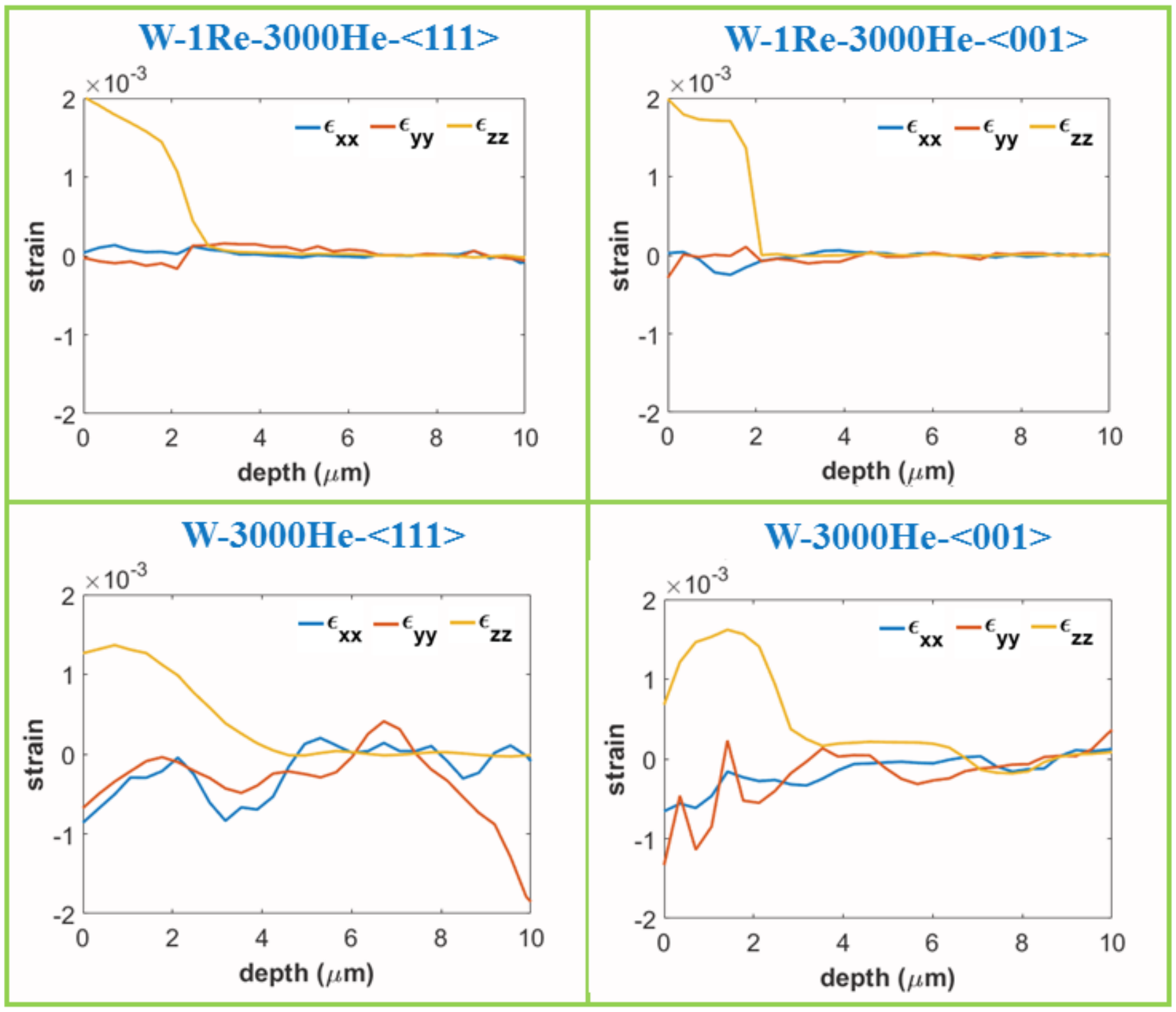

**Figure C.2** – Direct strains measured for samples implanted with 3000 appm of helium plotted vs sample depth.

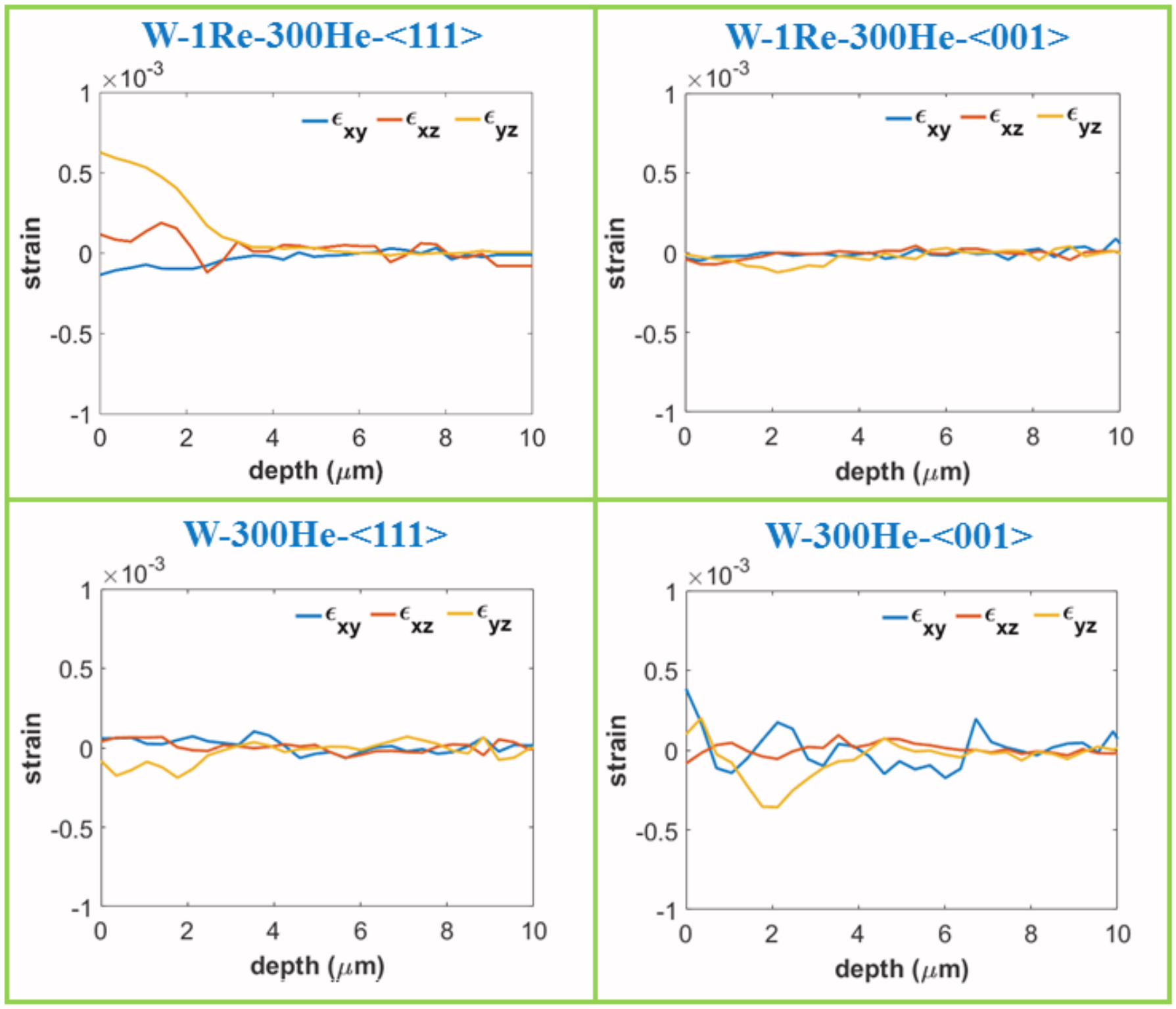

**Figure C.3** – Shear strains measured in samples implanted with 300 appm of helium plotted vs sample depth.

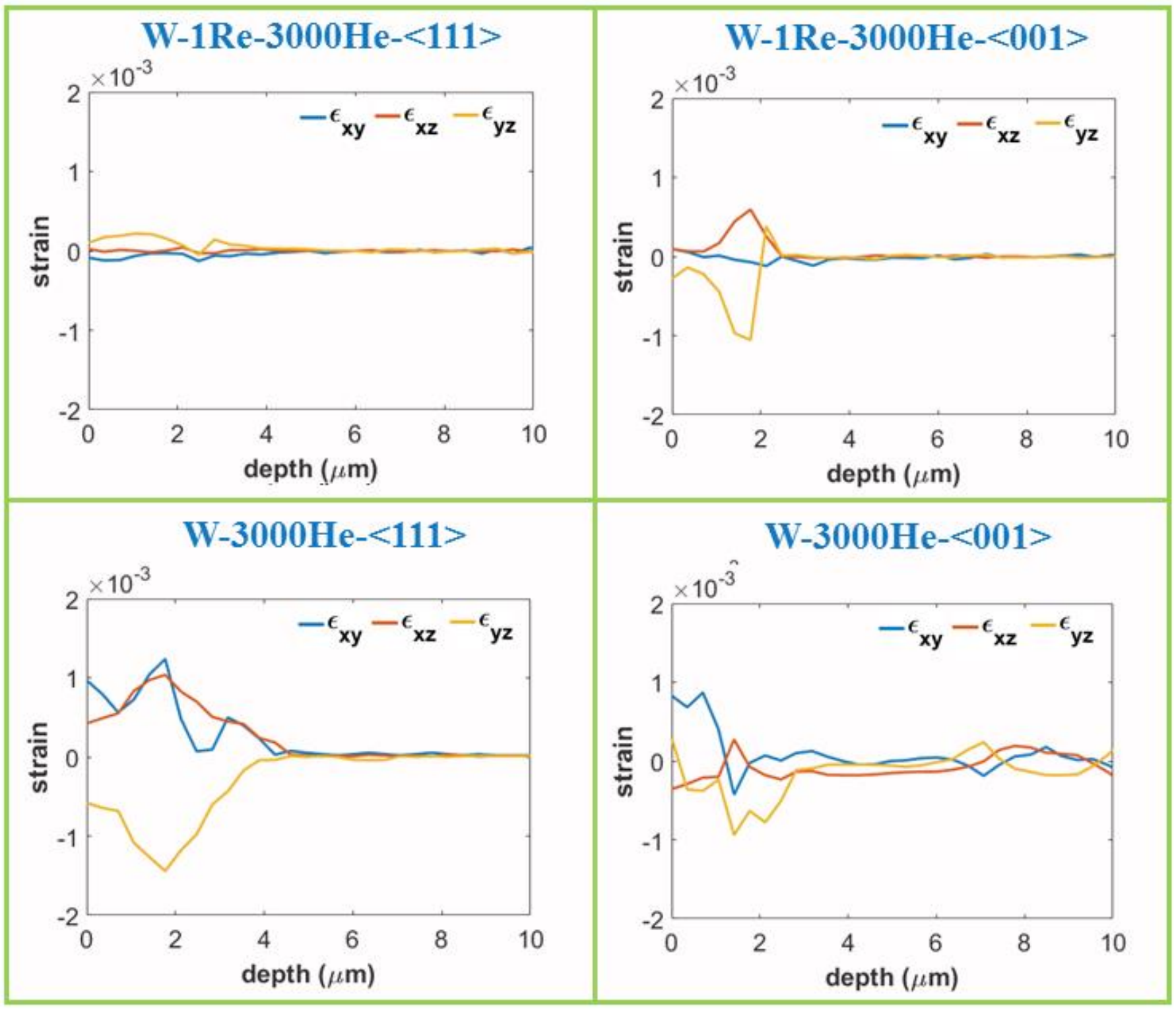

**Figure C.4** – Shear strains measured in samples implanted with 3000 appm of helium plotted vs sample depth.

# Appendix D

The derivation of Eq. (3) is given below.

Let the lattice strain measured experimentally be given by the tensor $\boldsymbol{\varepsilon}$ . The pure volumetric Eigenstrain is given by $\boldsymbol{\varepsilon}^*_{vol}$ and the correctional strain tensor term accounting for the lateral constraints on the implanted layer is given by $\boldsymbol{\varepsilon}_{corr}$. In this regime of small strain, $\boldsymbol{\varepsilon}$ may be additively decomposed as

$$\boldsymbol{\varepsilon} = \boldsymbol{\varepsilon}^*_{vol} + \boldsymbol{\varepsilon}_{corr} \tag{D.1}$$

The volumetric Eigenstrain tensor $\boldsymbol{\varepsilon}^*_{vol}$, is given by $\begin{bmatrix} \varepsilon^*_{vol}/3 & 0 & 0 \\ 0 & \varepsilon^*_{vol}/3 & 0 \\ 0 & 0 & \varepsilon^*_{vol}/3 \end{bmatrix}$. $\varepsilon_{corr_{11}} = \varepsilon_{corr_{22}} = -\varepsilon^*_{vol}/3$ due to the need for maintaining the geometrical continuity in the sample (shown by schematic in Figure D.1) and thus we assume $\varepsilon_{xx} = \varepsilon_{yy} \approx 0$. Considering Poisson effect

$$\varepsilon_{corr_{11}} = -\frac{\varepsilon^*_{vol}}{3} = \frac{1}{E}(\sigma_{corr_{11}} - \nu(\sigma_{corr_{22}} + \sigma_{corr_{33}})) \tag{D.2}$$

where, $E$ is the elastic modulus and $\nu$ the Poisson's ratio. $\sigma_{corr_{33}} = 0$ as there is a traction free boundary condition in the Z direction and $\sigma_{corr_{11}}$ is considered equal to $\sigma_{corr_{22}}$, owing to symmetry.

Thus,

$$\varepsilon_{corr_{11}} = -\frac{\varepsilon^*_{vol}}{3} = \frac{1}{E}(\sigma_{corr_{11}} - \nu(\sigma_{corr_{22}} + \sigma_{corr_{33}})) = \frac{\sigma_{corr_{11}}}{E}(1-\nu) \tag{D.3}$$

Rearranging we get,

$$\sigma_{corr_{11}} = \frac{-E * \varepsilon^*_{vol}}{3(1-\nu)} = \sigma_{corr_{22}} \tag{D.4}$$

Now $\varepsilon_{corr_{33}}$, may be written in terms of $\sigma_{corr_{11}}$ and $\sigma_{corr_{22}}$

$$\varepsilon_{corr_{33}} = \frac{1}{E}\left(\sigma_{corr_{33}} - \nu\left(\sigma_{corr_{11}} + \sigma_{corr_{22}}\right)\right) = \frac{2 * \varepsilon^*_{vol} * \nu}{3(1-\nu)} \tag{D.5}$$

As per Eq. (D.1), $\varepsilon_{33} = \varepsilon^*_{vol_{33}} + \varepsilon_{corr_{33}}$ i.e.

$$\varepsilon_{zz} = \frac{\varepsilon^*_{vol}}{3} + 2 * \left(\frac{\nu}{1-\nu}\right) * \left(\frac{\varepsilon^*_{vol}}{3}\right) \tag{D.6}$$

We note here, that the volumetric strain component of $\boldsymbol{\varepsilon}$, i.e. $\varepsilon_{vol}$ is given by $(\varepsilon_{xx} + \varepsilon_{yy} + \varepsilon_{zz})$ and is different from the volumetric Eigenstrain $\varepsilon^*_{vol}$. $\varepsilon_{vol} = \varepsilon_{zz}$ (as $\varepsilon_{xx} = \varepsilon_{yy} \approx 0$), would be equal to $\varepsilon^*_{vol}$, if $\nu = 0.5$. This does not apply for tungsten, which has $\nu = 0.28$.

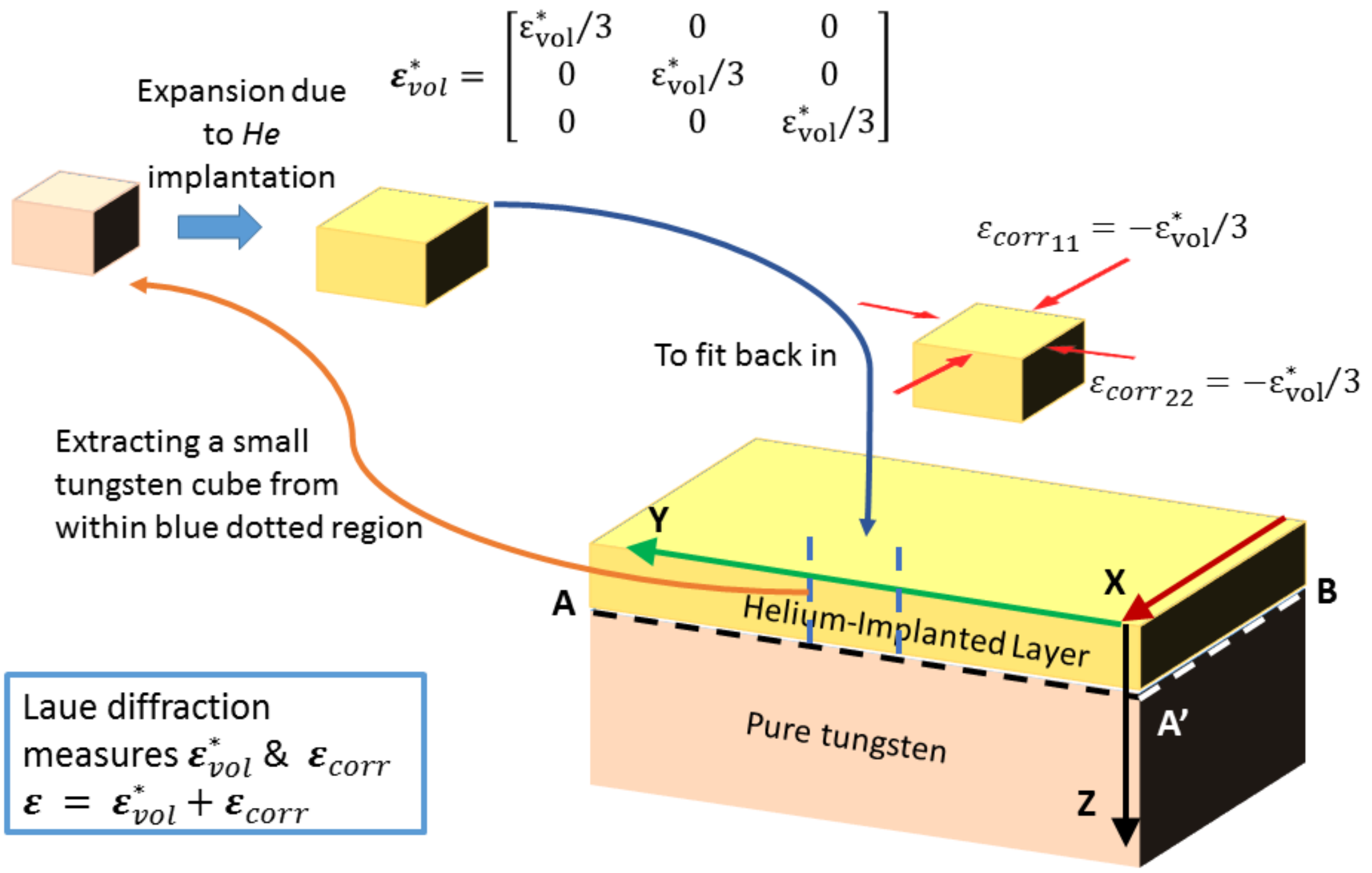

**Figure D.1 –** Schematic representation of the volumetric eigenstrain induced by helium-implantation and the correctional elastic strain generated in response to it.

## Data Availability

The raw data and MATLAB codes required to reproduce the findings for all the discussed samples are available to download from https://doi.org/10.5281/zenodo.1405136.

## List of Figures

## List of Tables